\pdfoutput=1
\documentclass[iop]{emulateapj}

\slugcomment{}

\shorttitle{Inner circumstellar environment of W42-MME MYSO}
\shortauthors{L.~K. Dewangan et al.}

\begin{document}

\title{Massive young stellar object W42-MME: The discovery of an infrared jet using VLT/NACO near-infrared images}
\author{L.~K. Dewangan\altaffilmark{1}, Y.~D. Mayya\altaffilmark{1}, A. Luna\altaffilmark{1}, and D.~K. Ojha\altaffilmark{2}}

\email{lokeshd@inaoep.mx}

\altaffiltext{1}{Instituto Nacional de Astrofísica, \'{O}ptica y Electr\'{o}nica, Luis Enrique Erro \# 1, Tonantzintla, Puebla, M\'{e}xico C.P. 72840.}
\altaffiltext{2}{Department of Astronomy and Astrophysics, Tata Institute of Fundamental Research, Homi Bhabha Road, Mumbai 400 005, India.}

\begin{abstract}
We report on the discovery of an infrared jet from a deeply embedded infrared counterpart of 6.7-GHz 
methanol maser emission (MME) in W42 (i.e. W42-MME). We also investigate that W42-MME drives a 
parsec-scale H$_{2}$ outflow, with detection of bow shock feature at $\sim$0.52~pc to the north. 
The inner $\sim$0.4~pc part of the H$_{2}$ outflow has a position angle of $\sim$18$\degr$ and 
the position angle of $\sim$40$\degr$ is found farther away on either side of outflow from W42-MME. 
W42-MME is detected at wavelengths longer than 2.2~$\mu$m and is a massive young 
stellar object, with the estimated stellar mass of 19$\pm$4~M$_{\odot}$.  
We map the inner circumstellar environment of W42-MME using VLT/NACO adaptive optics 
K$_{s}$ and L$^{\prime}$ observations at resolutions~$\sim$0\farcs2 and $\sim$0\farcs1, respectively.
We discover a collimated jet in the inner 4500~AU using the L$^{\prime}$ band, which contains
prominent Br$\alpha$ line emission.
The jet is located inside an envelope/cavity (extent $\sim$10640~AU) that is tapered at both ends
and is oriented along the north-south direction. Such observed morphology of outflow cavity around massive star is scarcely known and 
is very crucial for understanding the jet-outflow formation process in massive star formation.
Along the flow axis, which is parallel to the previously known magnetic field, two blobs are found in both the 
NACO images at distances of $\sim$11800~AU, located symmetrically from W42-MME. 
The observed W42-MME jet-outflow configuration can be used to constrain the jet launching and jet collimation models in massive star formation. 

%
\end{abstract}


\section{Introduction}
The driving mechanism of jets and their role in massive star formation (MSF) is still subject to considerable debate.
Outflows are an ubiquitous feature of MSF \citep{lada85,beuther02,zhang05}.
There have been increasing number of detection of molecular outflows from massive protostars 
\citep[e.g.][]{beuther02,beuther09,wu04,zhang05,arce07}, that suggests the accretion process for their formation.
However, the number of known massive young stellar objects (MYSOs) associated with collimated jets is 
very limited \citep[see][for more details]{guzman10}. 
Additionally, the information of magnetic field towards MYSO jet/outflow system is scarcely available \citep[e.g.][]{jones04,carrasco10}. 
The investigation of MSF is often handicapped by the fact that 
massive stars are statistically rare, situated at large distances ($\geq$~1 kpc), 
present in clustered environments, and suffer from high extinction. 
All these factors inhibit their observational study in the early phases. 
Key to overcome these difficulties is the high-angular resolution observations towards candidate 
massive protostars at infrared and longer wavelengths, which will allow mapping of the inner 
circumstellar environment. 
It has been well established that the 6.7-GHz methanol maser emission (MME) 
is associated with early phases of MSF \citep[e.g.][]{walsh98,urquhart13} and hence traces the locations of MYSOs.
Therefore, such sites are very promising to carry out a detailed study to understand the involved physical processes of MSF.

W42 is a Galactic giant H\,{\sc ii} region (G25.38$-$0.18) harboring a spectroscopically classified 
O5-O6 star in the heart of a near-infrared (NIR) cluster \citep{blum00}. 
Distances reported for W42 range from 2.2~kpc obtained by \citet{blum00} 
assuming the O-star is in the ZAMS, to 3.8~kpc obtained kinematically 
using radio recombination line H86$\alpha$ from the H\,{\sc ii} region \citep[59.1 km\,s$^{-1}$;][]{lester85}, 
and CO line from the associated cloud \citep[58--69~km\,s$^{-1}$;][]{anderson09}. 
Using the K-band polarimetric image towards the NIR cluster in W42, \citet{jones04} discovered 
a bipolar reflection nebula that is illuminated by an undetected source at $\sim$10$\arcsec$ south-west of the O-star. 
Based on the alignment of the long axis of the nebula with the magnetic field at the position angle of $\sim$15$\degr$, they argued that the reflection 
nebula is tracing a bipolar outflow driven by an embedded young stellar object (YSO) that is hidden by a cloud with extinction of A$_{V}$ $>$ 55 mag. 
However, they did not find any K-band counterpart of the driving source. 
We found that a 6.7~GHz MME detected by \citet{szymczak12} lies close to the predicted position of the driving source. 
We trace the reflection nebula as a parsec-scale bipolar outflow in the H$_{2}$ (1$-$0) S(1) 2.12 $\mu$m 
continuum-subtracted image (see Figure~\ref{fig1}). 
The line of sight velocity of the MME is $\sim$58.1 km s$^{-1}$, very similar to the 
velocity of the H\,{\sc ii} region. Considering the fact that the 6.7-GHz maser is physically 
associated with W42, a distance of 3.8 kpc is also adopted for MME in this paper. 
The presence of a parsec-scale outflow around 6.7-GHz MME makes this system ideal for 
exploring the early phases of MSF, including looking for a jet using high angular 
resolution NIR imaging.

In this paper, using multi-wavelength archival data, we investigate an infrared counterpart (IRc) of 6.7-GHz MME (hereafter W42-MME) and 
present observational evidence for W42-MME as a MYSO with a jet at the base of a parsec-scale bipolar H$_{2}$ outflow.
The presence of jet is investigated using Very Large Telescope (VLT) NIR adaptive-optics images and the stellar counterpart is investigated
using mid-infrared images.
\section{Data and analysis}
\label{sec:obser}
\subsection{Adaptive-optics near infrared imaging data}
Adaptive-optics imaging data towards 6.7-GHz MME were retrieved from the ESO-Science Archive 
Facility\footnote[1]{http://archive.eso.org/eso/eso\_archive\_main.html} (ESO proposal ID: 089.C-0455(A); PI: Joao Alves). 
The images were observed in K$_{s}$-band ($\lambda _{c}=2.18\, \mu \rm m, \Delta \lambda =0.35\, \mu \rm m$) and 
L$^{\prime}$-band ($\lambda _{c}=3.80\, \mu \rm m, \Delta \lambda =0.62\, \mu \rm m$), using 8.2m VLT with 
NAOS-CONICA (NACO) adaptive-optics 
system\footnote[2]{http://www.eso.org/sci/facilities/paranal/instruments/naco.html} \citep{lenzen03,rousset03}. 
Five K$_{s}$ frames and six L$^{\prime}$ frames of 24 and 21 seconds, respectively, were used in this study. 
These data were reduced using the standard analysis procedure available in IRAF and STAR-LINK packages, as described in detail by \citet{kumar13}. 
The astrometry of NACO images was calibrated using the GPS K-band point sources (see Section~\ref{subsec:data} for more details).
The plate scales of final processed NACO K$_{s}$ and L$^{\prime}$ images were 0\farcs054/pixel and 
0\farcs027/pixel, respectively, resulting in a resolution of 0\farcs2 ($\sim$760 AU) and 0\farcs1 ($\sim$380 AU),
respectively.
\subsection{Other Archival Data}
\label{subsec:data}
We also utilized the multi-wavelength data from different surveys (e.g. Multi-Array Galactic Plane 
Imaging Survey \citep[MAGPIS: 20 cm;][]{helfand06}, Galactic Ring Survey \citep[GRS: $^{13}$CO(J = 1$-$0);][]{jackson06}, 
APEX Telescope Large Area Survey of the Galaxy \citep[ATLASGAL: 870 $\mu$m;][]{schuller09}, 
{\it Herschel} Infrared Galactic Plane Survey \citep[Hi-GAL: 70--500 $\mu$m;][]{molinari10}, 
Galactic Legacy Infrared Mid-Plane Survey Extraordinaire \citep[GLIMPSE: 3.6--8.0 $\mu$m;][]{benjamin03}, 
UKIRT Wide-field Infrared Survey for H2 \citep[UWISH2: 2.12 $\mu$m;][]{froebrich11}, and 
UKIRT NIR Galactic Plane Survey \citep[GPS: 1.25--2.2 $\mu$m;][]{lawrence07}). 
To obtain a continuum-subtracted H$_{2}$ map (hereafter H$_{2}$ map), the GPS K-band image was 
registered and scaled to the UWISH2 narrow-band H$_{2}$ (1$-$0) S(1) 2.12 $\mu$m image (resolution $\sim$1$\arcsec$). 
The H$_{2}$ map was smoothed with a Gaussian of sigma $\sim$1 pixel to 
enhance the faint features. The K-band polarimetric data were obtained from \citet{jones04}. 
\section{Results}
\label{sec:data}
\subsection{Multi-wavelength view around 6.7-GHz MME}
\begin{figure}
\epsscale{1.1}
\plotone{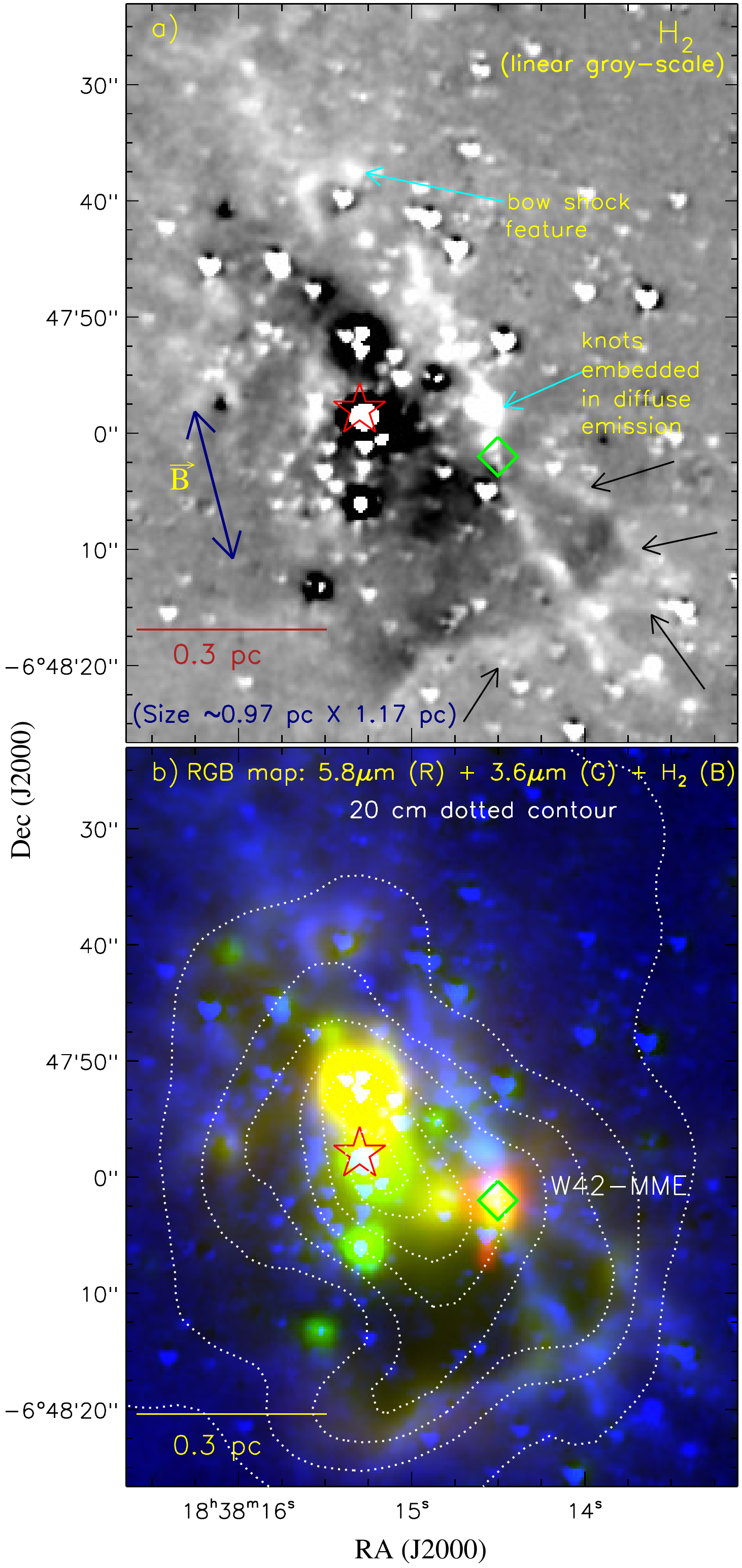}
\caption{\scriptsize a) Continuum-subtracted H$_{2}$ (1$-$0) S(1) 2.12 $\mu$m image illustrating the outflow.
b) RGB color image (in linear scale) with an overlay of the 20 cm continuum emission (0.211 Jy/beam $\times$ (0.1, 0.2, 0.3, 0.4, 0.55, 0.7, 0.85, 0.95)). 
In both the panels, the positions of a 6.7-GHz MME ($\Diamond$) and an O5-O6 star ($\star$) are marked in the figure. 
The 6.7-GHz MME appears at the center of a bipolar emission. 
The IRc coinciding with the 6.7-GHz MME (i.e. W42-MME) is clearly seen in the RGB image. 
The cyan arrows highlight noticeable features, including a bow shock, along a continuous elongated H$_{2}$ emission. 
Apart from the elongated H$_{2}$ emission, the black arrows also show H$_{2}$ emissions in the south-west and south directions. 
The long axis of the H$_{2}$ nebulosity is aligned with the magnetic field ($\vec{B}$) at the position 
angle of $\sim$15$\degr$ \citep[e.g.][]{jones04}. The magnetic field direction is also shown by a thick blue arrow. (Note: In the displayed H$_{2}$ map, 
the K-continuum is scaled so as to avoid negative values within the H\,{\sc ii} region, which has resulted in the under-subtraction of stellar features.)}
\label{fig1}
\end{figure}

The distribution of molecular H$_{2}$ emission is shown in Figure~\ref{fig1}a, 
where the most prominent extended structure is a continuous elongated emission 
feature centered on the position of the 6.7-GHz MME. 
On the northern side, the feature intercepts a bow shock at a projected
distance of $\sim$0.52~pc from the 6.7-GHz MME.  
The explanation of the origin of H$_{2}$ emission in W42 is difficult 
because an O5-O6 star is located at $\sim$0.22~pc distance from the MME in W42 H\,{\sc ii} region.
The presence of H$_{2}$ emission can be interpreted by either ultraviolet (UV) fluorescence or shocks. 
Considering the observed morphology of the H$_{2}$ features, 
we suggest that the elongated H$_{2}$ emission is caused by shock, which traces 
a parsec-scale bipolar H$_{2}$ outflow. The presence of bow shock feature and knots 
embedded in diffuse emission along the elongated H$_{2}$ feature offers further evidence for the shock-excited outflow activity. 
The overall morphology of the H$_{2}$ feature resembles the bipolar outflows of Herbig Haro objects. 
The outflow nature of the observed H$_{2}$ feature is further supported by 
recent detection of [Fe~II] emission coinciding with the southwest part 
\citep[see Figure~12 of][]{lee14}.
The inner $\sim$0.4~pc part of the H$_{2}$ outflow has a position angle of $\sim$18$\degr$, which coincides with the reflection nebula 
reported by \citet{jones04} (see their Figure~5).  However, the H$_{2}$ feature continues 
much further than the reflection nebula on either side at a position angle of $\sim$40$\degr$. 
Considering the variation in position angles along the outflow, the outflow morphology appears a little twisted in ``S" shape. 
The GRS $^{13}$CO(J = 1$-$0) line data are unable to provide any outflow signatures towards MME due to coarse beam (beam size $\sim$45$\arcsec$).
Additionally, there are no high-resolution CO observations available in the literature.

\begin{figure}
\epsscale{1.2}
\plotone{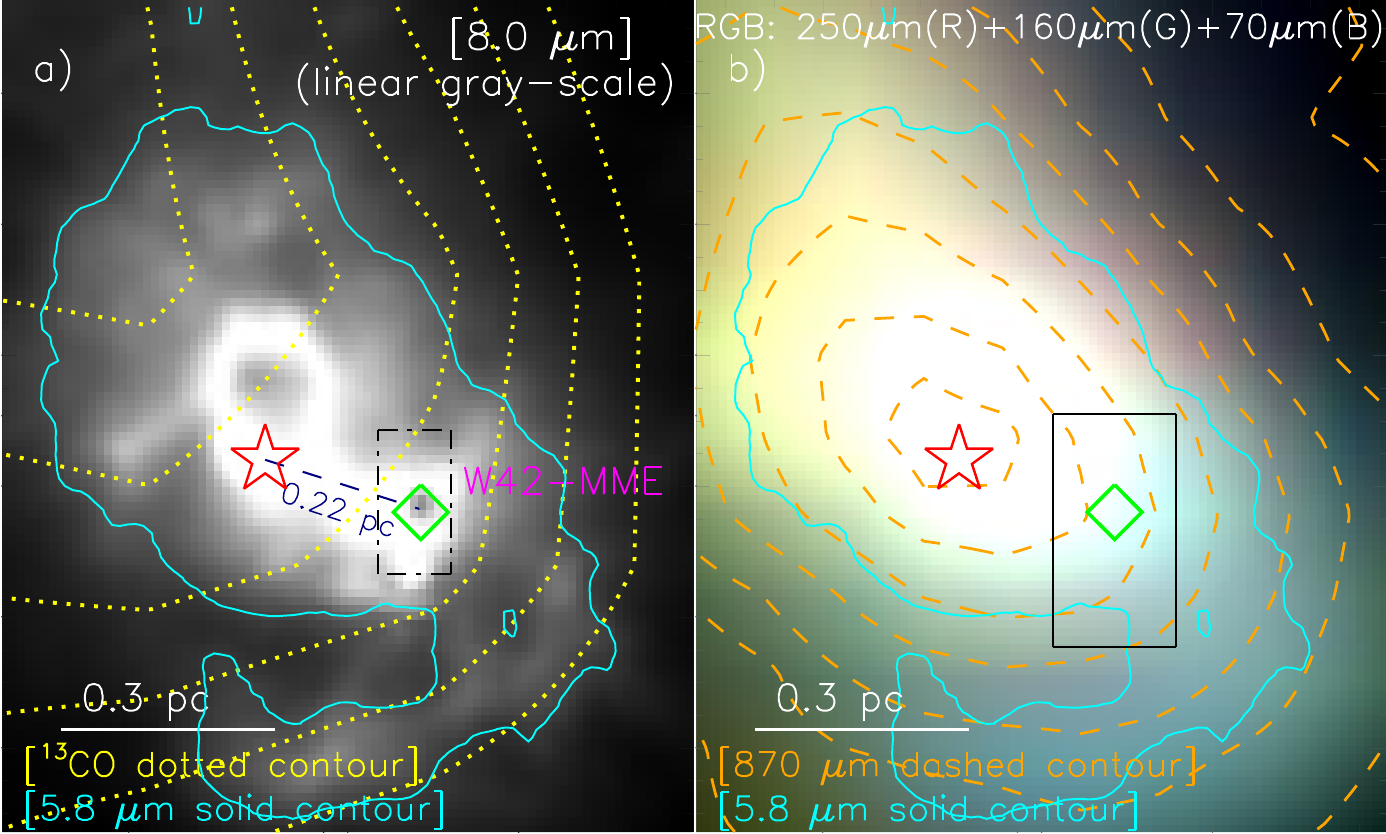}
\caption{\scriptsize a) Overlay of the GRS $^{13}$CO dotted contours on the 8.0 $\mu$m 
image. The dot-dashed black box is shown as a zoomed-in view in Figure~\ref{fig4}. b) Color composite map using {\it Herschel} images (in logarithmic scale). The ATLASGAL 870 $\mu$m emissions are overlaid by orange contours 
with 10, 20, 30, 40, 55, 70, 85, and 95\% of the peak value (i.e., 3.3 Jy/beam).
The solid black box is shown as a zoomed-in view in Figure~\ref{fig5}a. 
In both the panels, the other marked symbols are similar to those shown in Figure~\ref{fig1}. 
The boundary of a cavity-like structure is shown by the 5.8 $\mu$m cyan contour with a level of 555 MJy/sr in both the panels.} 
\label{fig2}
\end{figure}

In order to identify the IRc of 6.7-GHz MME, 
we present H$_{2}$, 3.6 $\mu$m, and 5.8 $\mu$m images in Figure~\ref{fig1}b. 
W42-MME is barely detected in the GPS K-band image (not shown here) with upper limit of $\sim$12.815 mag. 
It is well detected in GLIMPSE 3.6--5.8 $\mu$m images but saturated in the 8.0 $\mu$m (see Figures~\ref{fig1}b and~\ref{fig2}a). 
The GLIMPSE-I catalog lists photometric magnitude only in one of the GLIMPSE bands. 
Therefore, aperture photometry was carried out for W42-MME using the GLIMPSE images \citep[see][for more details]{dewangan15}. 
The GLIMPSE photometric magnitudes were measured to be 8.295$\pm$0.031 (3.6 $\mu$m), 6.178$\pm$0.012 (4.5 $\mu$m), 
and 4.545$\pm$0.103 (5.8 $\mu$m) mag. The color excess criteria further reveal this source to be a Class I YSO 
\citep[see][for classification schemes]{dewangan15}. 
Figure~\ref{fig1}b also illustrates the distribution of ionized emission using MAGPIS 20~cm map (beam size $\sim$6$\arcsec$). 
It can be noticed that the MME-outflow system is immersed within the W42 H\,{\sc ii} region. 
The GRS $^{13}$CO(J = 1$-$0) molecular, ATLASGAL dust continuum at 870~$\mu$m (beam size $\sim$ 19$\farcs$2), 
and {\it Herschel} sub-millimetre emissions (beam size $\sim$6--18$\arcsec$) are also peaked close to the 6.7-GHz MME position (see Figure~\ref{fig2}). 
These data therefore suggest that W42-MME is located at the densest part of the molecular cloud in W42. 

Apart from the elongated bipolar feature, the H$_{2}$ map also shows H$_{2}$ emissions in the south-west and south directions, as highlighted by 
black arrows (see Figure~\ref{fig1}a). These particular H$_{2}$ features are nicely coincident with GLIMPSE infrared and 20 cm emissions. 
In general, GLIMPSE images (3.6--8.0 $\mu$m) have been used to trace 
a photodissociation region around the H\,{\sc ii} region through the presence of polycyclic 
aromatic hydrocarbon features at 3.3, 6.2, 7.7, and 8.6 $\mu$m. 
Therefore, it seems that these H$_{2}$ features (see black arrows in Figure~\ref{fig1}a) are formed 
due to UV fluorescence, tracing the wall of an ionized cavity-like structure (see Figure~\ref{fig1}b). 
The ionized cavity-like structure seen in GLIMPSE images is also shown in Figure~\ref{fig2}. 
Interestingly, the bipolar H$_{2}$ feature in the south direction is further extended beyond the cavity-like structure (Figure~\ref{fig1}). 

\begin{figure}
\epsscale{1.2}
\plotone{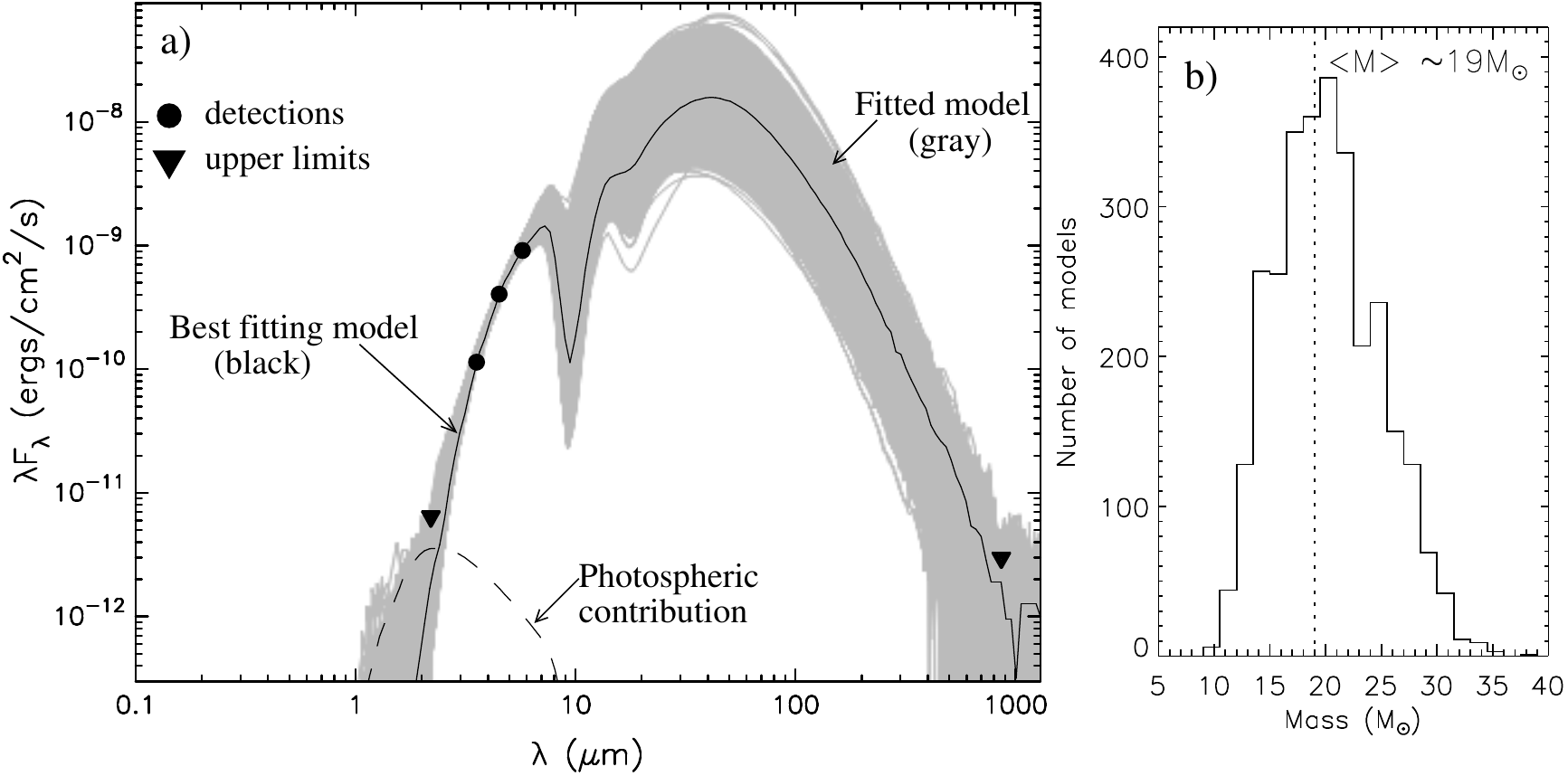}
\caption{\scriptsize a) The observed SED (filled symbols) for W42-MME, obtained with the \citet{robitaille06} 
fitting procedure. 
Only those models are selected that satisfy the condition ($\chi^{2}$ - $\chi^{2}_{best}$) per data point $<$ 3. 
The black solid line corresponds to the best fit; the gray solid curves represent all other models providing a good fit to the data. 
The dashed curve shows photospheric contribution. b) Distribution of the stellar mass for all plotted models in Figure~\ref{fig3}a. 
We derive a weighted mean mass of 19$\pm$4 M$_{\odot}$.} 
\label{fig3} 
\end{figure}

To estimate the stellar mass, we modelled the Spectral Energy Distribution (SED) of W42-MME using an on-line SED modelling tool \citep{robitaille06,robitaille07}. 
{\it Spitzer}-GLIMPSE fluxes at 3.6, 4.5, and 5.8 $\mu$m bands were used as data points, 
whereas GPS K-band magnitude and ATLASGAL 870 $\mu$m flux \citep[obtained from][]{csengeri14} were treated as upper 
limits \citep[e.g.][]{dewangan15}. 
We provided the visual extinction in the range 0--70 mag and our adopted distance to W42-MME, as input parameters for SED modelling. 
The fitted SED models of W42-MME are shown in Figure~\ref{fig3}a. 
The stellar mass distribution for W42-MME is also plotted in Figure~\ref{fig3}b. 
The weighted mean values of the stellar mass and extinction are 19$\pm$4 M$_{\odot}$ and 48$\pm$15 mag, respectively. 
The extinction value of W42-MME is consistent with the estimates of \citet{jones04}. 
The weighted mean value of age is found to be 8.1 $\times$ 10$^{4}$ year. 
Our SED result suggests that W42-MME is a MYSO candidate with a weighted mean luminosity of 4.5 $\times$ 10$^{4}$ L$_{\odot}$. 
We also fitted the SED assuming a distance of 2.2 kpc and estimated a weighted mean mass of 14$\pm$3 M$_{\odot}$. 
W42-MME is still a massive protostar candidate at a shorter distance of 2.2 kpc. 
The SED result of W42-MME is also in agreement with the generally accepted argument that the 6.7-GHz methanol masers 
are solely associated with MSF \citep[e.g.][]{urquhart13}. 

Taken together, it appears that W42-MME is a driving source of a parsec-scale bipolar outflow and 
the axis of the outflow is parallel to the magnetic field (see Figure~\ref{fig1}). 
\subsection{Inner circumstellar environment of W42-MME}
\begin{figure}
\epsscale{1.2}
\plotone{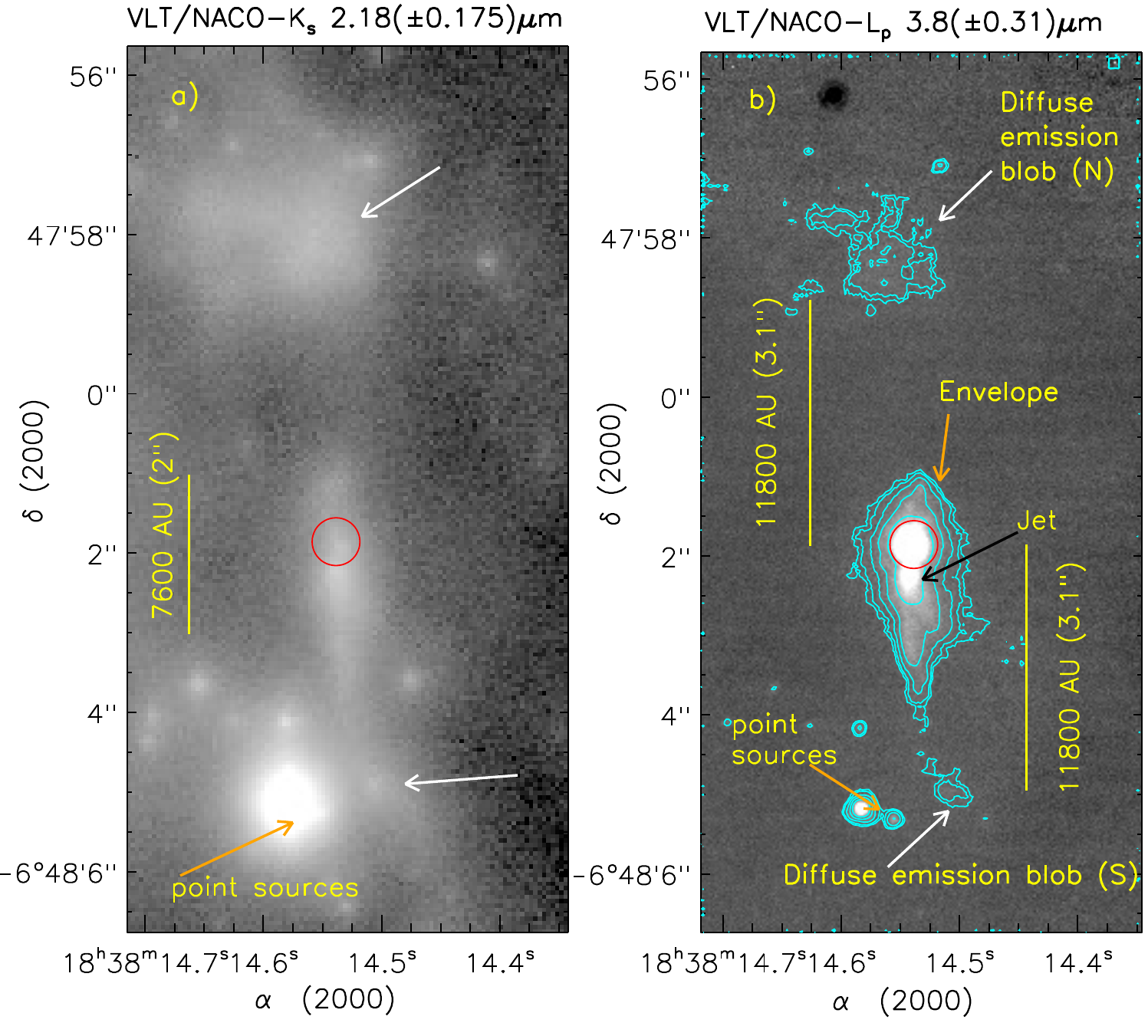}
\caption{\scriptsize VLT/NACO adaptive-optics images  (in logarithmic gray scale) of W42-MME. 
The selected area of NACO images is shown by a box in Figure~\ref{fig2}a. 
a) K$_{s}$ band. Diffuse emissions are highlighted by white arrows (also see Figure~\ref{fig4}b). 
b) Overlay of L$^{\prime}$ contours on the L$^{\prime}$ image (the outer most contour corresponds to 4$\sigma$, 
with successive contours increasing logarithmically). 
In both the panels, the position of the driving source 
($\alpha_{2000}$ = 18$^{h}$38$^{m}$14$^{s}$.54, $\delta_{2000}$ = $-$06$\degr$48$\arcmin$01$\arcsec$.86) 
is shown by a red circle. Distances of diffuse emissions seen in L$^{\prime}$ image are also highlighted with respect to W42-MME.}
\label{fig4}
\end{figure}

In Figure~\ref{fig4}, we show the inner circumstellar environment of W42-MME using 
the VLT/NACO adaptive-optics images at K$_{s}$ and L$^{\prime}$ bands. 
The K$_{s}$ image shows diffuse emissions near the position of 6.7-GHz MME, 
as well as, in north and south directions (see arrows in Figure~\ref{fig4}a), which are coincident with the emissions seen in H$_{2}$ map. 
The L$^{\prime}$ image reveals a point source (i.e. W42-MME) associated with a small-scale feature (within a scale of $\sim$4500 AU). 
The small-scale feature is noticeably elongated in vertical direction and the full width half maximum of its point spread function is almost twice that of 
other stars seen in L$^{\prime}$ image.
This comparison allows us to rule out the small-scale feature as a point-like source. 
W42-MME is also surrounded by an envelope-like emission, which is extended on a physical scale of $\sim$10640 AU, oriented 
along the north-south direction.

\citet{zhang11} performed the radiative transfer calculations to understand the observed NIR and mid-infrared emissions around 
embedded MYSOs. They suggest that the outflow cavity is the most significant features in images up to 70 $\mu$m.
They also find that the NIR emission is due mainly to scattered light from the outflow cavity wall. 
Using the prediction of model, we consider the envelope-like structure as an outflow cavity traced by the scattered light. 
The cavity has very narrow opening tip at both ends and is more extended ($\sim$4900 AU in east-west) 
near the position of W42-MME (see Figure~\ref{fig4}b). 

\begin{figure}
\epsscale{1.2}
\plotone{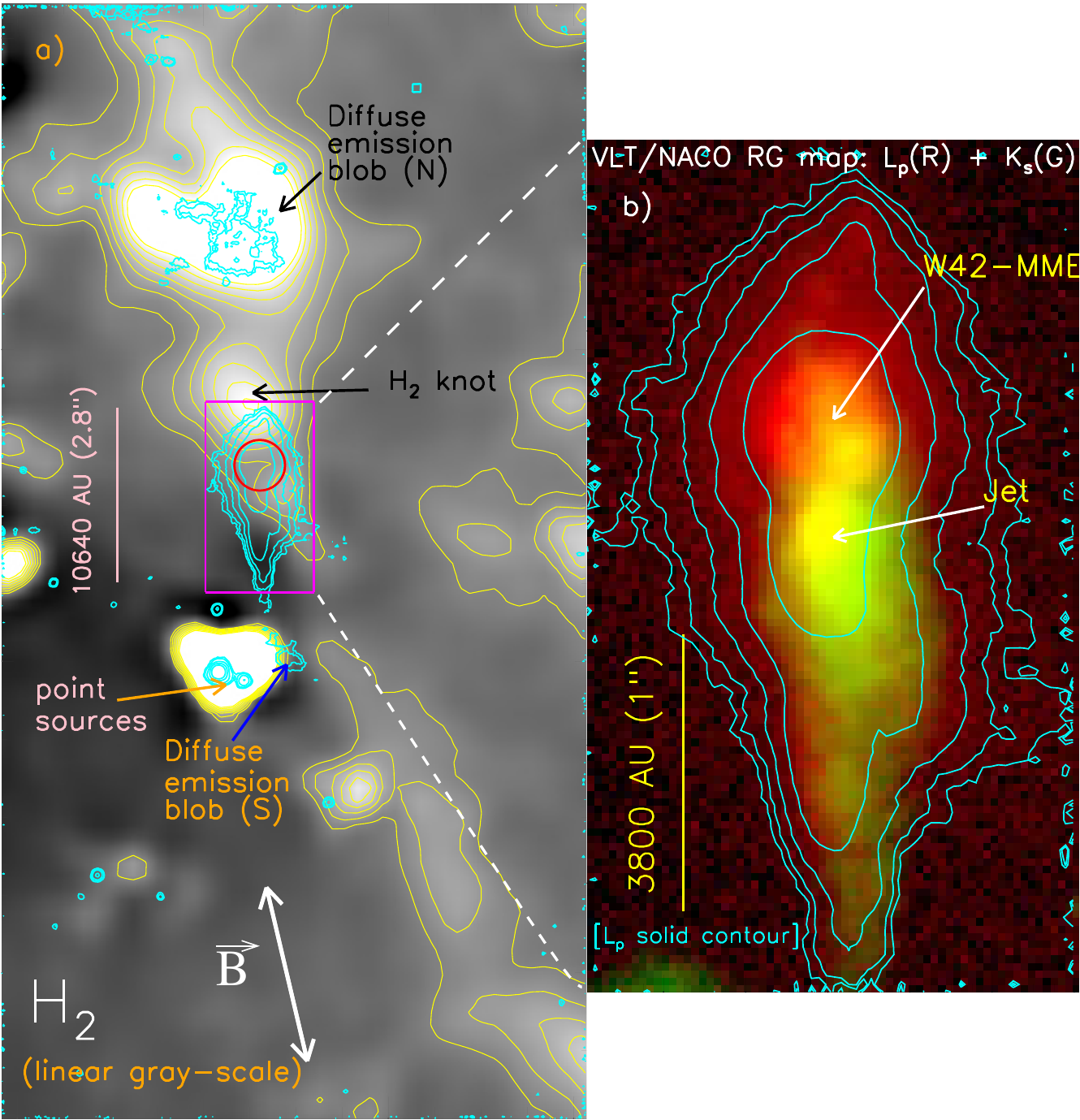}
\caption{\scriptsize a) Overlay of L$^{\prime}$ contours (in cyan color) on the continuum-subtracted H$_{2}$ image at 2.12 $\mu$m. 
The selected area of H$_{2}$ image is shown by a box in Figure~\ref{fig2}b. 
The L$^{\prime}$ contours are similar to those shown in Figure~\ref{fig4}b. 
The H$_{2}$ contours (in yellow) are superimposed in order to highlight faint H$_{2}$ features. 
The magnetic field ($\vec{B}$) direction is also shown as Figure~\ref{fig1}a. 
b) Zoomed-in view of a two color composite NACO image (in logarithmic gray scale), which clearly illustrates the morphology of cavity and jet.} 
\label{fig5} 
\end{figure}

In both NACO images, the diffuse emissions are found parallel to the cavity in north and south directions, 
which are separated by a similar distance $\sim$3\farcs1 ($\sim$11800 AU) with respect to W42-MME. 
It provides the evidence for the presence of a symmetric configuration along flow axis. 
The physical association of small-scale feature, the cavity, and the outflow is 
shown in Figure~\ref{fig5}a. In the north direction, the H$_{2}$  knot is coincident with the tip of the cavity along the flow axis. 
All these observed features suggest that they can be originated by the interaction 
of a jet with molecular gas in its surroundings. Therefore, the small-scale feature within the cavity is 
witness of a jet associated with MYSO W42-MME (see Figure~\ref{fig5}b for zoomed-in view). 

Considering together the observed emissions, 
NACO L$^{\prime}$ image traces the inner part of the outflow, which is highly collimated to a small-angle jet.
A zoomed-in view of the cavity and the jet using a two color composite NACO image (red:L$^{\prime}$ and green:K$_{s}$) is shown in Figure~\ref{fig5}b. 
The IRc and the cavity have red colors suggesting absence of K$_{s}$ emission from the parts.
It also appears that the cavity resembles an onion-like structure. 
The jet is bright in L$^{\prime}$ image, which encompasses the Br$\alpha$ (4.05 $\mu$m) line emission as main contributor.
Therefore, we suggest that it is an ionized jet associated with a driving source W42-MME. 
The K$_{s}$ diffuse emission is also coincident with jet-like feature located within the cavity, which 
could be due to the Br$\gamma$ (2.166 $\mu$m) line emission. 
In order to further investigate the ionized nature of the jet, we estimated the emission measure (EM) of jet feature 
using its photometric fluxes obtained in both the NACO images. The photometric calculations of jet-feature were 
performed with the removal of its surrounding diffuse emission. 
In the calculation of EM, we assume that the whole jet emission is dominated by the hydrogen lines.
Therefore the K$_{s}$ and L$^{\prime}$ fluxes of jet feature provide a measure of the Br$\gamma$ and Br$\alpha$ 
line emissions, respectively.
Following the recombination theory and the observed flux ratio of Br$\alpha$ to Br$\gamma$, we computed the 
extinction at Br$\alpha$ and Br$\gamma$ emissions \citep[see Appendix to][and references therein]{ho90}. 
The intrinsic flux ratio of Br$\alpha$ to Br$\gamma$ was adopted for Case~B with T$_{e}$=10$^{4}$ K and n$_{e}$=10$^{4}$ cm$^{-3}$. 
Following \citet{indebetouw05} extinction law, the visual extinction (A$_{V}$) of jet emission is also estimated about 37 
mag using the extinction at Br$\gamma$ emission,  which is in agreement with our SED result.
Furthermore, the extinction corrected Br$\gamma$ flux is used to measure the EM of $\sim$2.7 $\times$ 10$^{9}$ cm$^{-6}$ pc 
\citep[see][for the equation]{tokunaga79}, which can be taken as an indicative value. 
Our estimate of EM is consistent with the values obtained for candidates radio jets toward MYSOs \citep[e.g.][]{guzman12}. 
It is also to be noted that the magnetic field is parallel to the jet-outflow axis (see Figure~\ref{fig5}a). 
In the south direction, the blob emissions and the tip of the cavity are not perfectly aligned with the H$_{2}$ outflow axis.  
Furthermore, the H$_{2}$ outflow has a little bend with respect to the jet axis. 
It further suggests the variation in position angles from small scale to large scale along the outflow axis.  
\subsection{Summary and Discussion}
We studied the inner circumstellar environment of a deeply embedded IRc of 6.7-GHz MME in W42 (i.e. W42-MME), 
as the driving source of a parsec-scale bipolar H$_{2}$ outflow. 
The parsec-scale H$_{2}$ outflow morphology appears a little twisted in ``S" shape, with the position angle changing smoothly from 
$\sim$18$\degr$ near the IRc to $\sim$40$\degr$ at the ends. 
W42-MME is characterized as a MYSO with mass of 19$\pm$4 M$_{\odot}$. 
This is the first time we identified W42-MME and characterized its stellar mass. 
The aim of this work was to understand the jet/outflow formation process in MSF. 
We discovered a collimated jet in the inner 4500~AU in W42-MME using the L$^{\prime}$ band to which Br$\alpha$ line emission 
contributes prominently, suggesting the jet is ionized in nature. 
The presence of jet was further confirmed by the detection of bow shock feature and knots in diffuse emission in H$_{2}$ map 
along the flow axis (i.e. jet-related features). 
The jet is found inside a cavity (extent $\sim$10640~AU) oriented along the north-south direction, 
which is tapered at both ends. 
Such a morphology appears to be responsible for collimating the jet/outflow.  
The  diffuse nebular blobs seen in north and south directions are located at similar distances ($\sim$11800~AU) from W42-MME, 
which can be considered as a symmetric configuration along the flow axis. 
Previously published magnetic field direction is parallel to the jet-outflow flow axis. 

All these observed characteristics at a few thousands AU and a parsec-scale share some common features of the jet-driven bow shock model, 
the precessing jet model, and the magnetically-driven model \citep[e.g.][and references therein]{arce07,pudritz07}.
In summary, a detailed modelling is needed to pinpoint the exact physical mechanism for W42-MME jet-outflow system.

The high angular resolution of NACO images around a MYSO W42-MME have enabled the detection of the inner jet-outflow configuration. 
Such a system is hardly known and is very crucial for understanding the jet-outflow formation process. 
We find that the morphology of outflow cavity in W42-MME is unique and rare, which is not seen at the 
inner environment of some well-studied low and massive stars \citep[e.g.][]{valusamy07,fuller01,carrasco10}. 
Therefore, the inferred spatial morphology of W42-MME can be used to constrain the jet launching 
and jet collimation models in MSF. 
\acknowledgments
LKD is supported by the grant CB-2010-01-155142-G3, from the CONACYT (M\'{e}xico). 
The research is supported by CONACYT (M\'{e}xico) grants CB-2010-01-155142-G3 (PI. YDM) and CB-2012-01-1828-41 (PI. AL). 
We thank the anonymous referee for the useful comments which improved this paper. 
We also thank Dr. M. S. Nanda Kumar for useful discussion.

\end{document}